\documentclass[final,3p,times]{elsarticle}

\usepackage{amssymb}
\usepackage{hyperref} 
\usepackage{CJKutf8} 
\usepackage{amsmath} 
\usepackage{xcolor} 
\usepackage{pdfpages} 

\begin{document}
\begin{frontmatter}



\title{``Born in Rome" or ``Sleeping Beauty": Emergence of hashtag popularity on the Chinese microblog Sina Weibo}


\author[ceu]{Hao Cui}
\author[ceu]{János Kertész} 
\ead{kerteszj@ceu.edu}
\cortext[cor1]{Corresponding author}

\address[ceu]{{Department of Network and Data Science, Central European University},
	            {Quellenstrasse 51}, 
            {Vienna},
            { A-1100}, 
            {Austria}}

\begin{abstract}
To understand the emergence of hashtag popularity in online social networking complex systems, we study the largest Chinese microblogging site Sina Weibo, which has a Hot Search List (HSL) showing in real time the ranking of the 50 most popular hashtags based on search activity. We investigate the prehistory of successful hashtags from 17 July 2020 to 17 September 2020 by mapping out the related interaction network preceding the selection to HSL. We have found that the circadian activity pattern has an impact on the time needed to get to the HSL. When analyzing this time we distinguish two extreme categories: a) ``Born in Rome", which means hashtags are mostly first created by superhubs or reach superhubs at an early stage during their propagation and thus gain immediate wide attention from the broad public, and b) ``Sleeping Beauty", meaning the hashtags gain little attention at the beginning and reach system-wide popularity after a considerable time lag. The evolution of the repost networks of successful hashtags before getting to the HSL show two types of growth patterns: ``smooth" and ``stepwise". The former is usually dominated by a superhub and the latter results from consecutive waves of contributions of smaller hubs. The repost networks of unsuccessful hashtags exhibit a simple evolution pattern.

\end{abstract}





\end{frontmatter}


\section{Introduction}
Microblogging sites are online social networking platforms where users interact with each other through activities such as post, repost, comment, reply, mention or like. When it comes to the definition of popularity on social media, researchers had various metrics by regarding popularity of an online content as the frequency of daily occurrence\cite{zhang2016creates}, the number of reposts/views at a time\cite{bao2013popularity, ma2013towards}, the cascade size\cite{goel2016structural}, or straightforwardly the displayed list of popular items by social media platform providers. Users on microblogging sites generate hundreds of millions of posts per day, some of which contain one or multiple hashtags referring to the topic of the posts. Among the flooding information, hashtags achieve different levels of popularity at a certain time, and the most popular hashtags in the whole system are depicted in ranking lists to inform users. Twitter Trends and Sina Weibo Hot Search List (HSL) are examples of microblogging ranking lists which show in real time the most popular hashtags that gain wide attention in the whole microblogging system. These lists serve as indication of public interest and attention~\cite{Annamoradnejad2019twitter_trend,ch_jk} and, at the same time, they trigger collective awareness~\cite{asur_2011} of the latest trending topics or events emerging in the world. These trends or hot topics can originate from natural reaction to real-world events~\cite{Thij_2014} or from manipulation by companies and organizations~\cite{Ratkiewicz_2011}.  

Twitter is a worldwide service, with 217 million monetizable daily active users \cite{twitter} as of the fourth quarter of 2021. Its Chinese counterpart, Sina Weibo, the most popular microblogging service in China, has 248 million daily active users and 573 million monthly active users~\cite{mau} as of the third quarter of 2021, who generate and propagate information in the whole Weibo system. Accordingly, Sina Weibo has become a popular tool for the Chinese public to look for information. The topics covered by hashtags which occur on the HSL are very diverse, to name a few, social events, TV programs, celebrities, entertainment, health and politics. During the time of COVID-19, the Weibo HSL played an important role in keeping people aware of the COVID-related news and updates~\cite{ch_jk}. Since the HSL has an advertising effect on the hashtags to the public, it is natural that celebrities or companies would desire a position on the HSL or disappear from the HSL in the case of a negative influence~\cite{jiang_fan}. Studies have also identified online censorship~\cite{censor, censor2} control practices and the possibility of algorithmic intervention~\cite{ch_jk} on Sina Weibo.

It is of primary interest to understand why some hashtags get to the top rank list and others not. Understanding the emergence of these popular topics plays important role in marketing, governance and trend predictions in real world~\cite{Ma_2013, Thij_2014}. Researchers on Twitter information diffusion have observed variation in the spread of online information across topics~\cite{Romero_2011}, shown the effects of the content~\cite{Tsur_2012}, semantic characterization~\cite{Lehmann_2012} and the co-occurrence of hashtags~\cite{Pervin_2015} on popularity. Approaches have been introduced to predict the popularity peaks of the new hashtags on Twitter  from the perspectives of machine learning~\cite{Ma_2013_prediction}, and taking the context of Twitter social network \cite{Yu_2020}. Previous study on Twitter trending topics has shown retweets by other users are more important than the number of followers in determining trends~\cite{asur_2011}. Some studies modeled user behaviors to capture the emergence of Twitter trending topics based on characteristics of the retweet graphs~\cite{Thij_2014}. A recent study on Twitter trending topics has investigated real-time Twitter Trends detection along with the ranking of the top terms~\cite{Khan_2021}.

Although Sina Weibo has received less academic attention than Twitter, while it has overtaken Twitter in terms of the number of users, research on its popularity appeared soon after its launch.  The attention of the researchers studying popularity in Sina Weibo turned gradually to the HSL, earlier they have presented evolution analysis of trending topics~\cite{yu2012artificial}, long-term variation of popularity~\cite{zhang2016creates}, prediction of hot topics based on content quality and structural characteristics of early adopters~\cite{bao2013popularity}, as well as bursty human activity patterns~\cite{wu2022revealing}.

Why and how does some online information become popular are important questions. While many studies have been focusing on the ``why" part, putting efforts in identification of factors that lead to information popularity, less attention has been paid to what the patterns of the temporal network dynamics look like in the history of these contents before they achieve a high level of popularity measured in our case by getting to the HSL. Different routes to popularity have been observed for memes. Studies on information diffusion have found pervasive existence of two consecutive spikes of popularity in the diffusion of different information from different media during the whole lifetime of a meme's propagation \cite{zhang2017sleeping}. Studies on tweet's popularity have shown tipping points \cite{ma2013towards} may emerge through the lifecycle. What makes the history of those popular hashtags specific? What mechanisms can be deduced from the network dynamics in the early stage? In our study, we focus particularly on the popularity emergence period, which is the prehistory \emph{before} a hashtag becomes popular enough to appear on the HSL.

Our goal is to unfold the different routes of hashtags leading them to the HSL by studying the repost networks as well as their giant components during the time period from the first creation of the hashtags to their first appearance on the HSL. We investigate the influencing factors of the time needed for a successful hashtag to get to the HSL and identify the time of the day when the hashtag was born and the effect of huge hubs. The evolution of the repost network show either smooth or stepwise character which are also related to the above factors.

The paper is organized as follows: In the next section we describe the methods and data, followed by the results of our investigations. The paper terminates with a discussion section.
A Supplementary Information with videos on the repost network evolution complements the paper. 

\section{Methods}
\subsection{Data description} 

Sina Weibo is the most popular microblogging site in mainland China~\cite{Zahng:2017}, where Twitter and other Western online social media are blocked. It works similarly to Twitter as users may follow others and have followers, they can post texts and pictures, add hashtags to them, react to others' posts, and repost them. Sina Weibo is a major vehicle of self expression especially for young Chinese people and a forum for social movements~\cite{BBC}.

The popularity of hashtags on Sina Weibo emerges as users participate in the search for them, in the discussion on them and in their spreading. Like other microblogging sites, Sina Weibo also creates a real time ranking of the 50 most popular hashtags to inform users. As the name of the ranking list indicates, the Hot Search List (HSL) is based to a large extent on the search activity related to the hashtags, however, the concrete algorithm has been unknown and has been target of criticism. The latter can be understood as getting to the HSL not only informs users about popularity but also boosts it a lot, which, in turn, may have severe financial consequences. 
As a response to the criticism, on 23 August 2021, Sina Weibo released what it called the rule of capturing the ``hotness" $H$ of a hashtag at a certain time~\cite{weibo_announce}. The corresponding formula is as follows:
\begin{equation}
    H=(S_H+D_H+R_H)\times I_H,
\end{equation}
where $S_H$ is search hotness, referring to the search volume, including manual input search and click-and-jump search, $D_H$ refers to the amount of discussion, including original posting and re-posting, $R_H$ is the volume of readings
in the spreading process of the hashtag, and $I_H$ refers to the interaction rate of hot search results page. While Sina Weibo emphasizes the objectiveness and fairness of HSL, it admits at the same time to ``promote positive content" and that ``official media reports shall prevail" in case of major negative social events~\cite{weibo_announce} and the intervention in other cases (redundancy, serious inaccurate information as identified by government departments of content inducing severe conflicts). In fact, indication of intervention has been noticed on the HSL by statistical analysis~\cite{ch_jk}. 
{There are also commercial hashtags that are paid for the positions on the HSL, which often occupy the third or/and fifth ranks and marked with ``Recommendation (\begin{CJK*}{UTF8}{gbsn}荐\end{CJK*})", we don't consider their popularity emergence.
In 2021, Weibo changed the mark of the commercial hashtags, 
in similar positions while not anymore associated with a rank in the front. This study excludes those commercial hashtags.} 

We wrote a web scraper to crawl Sina Weibo HSL from 17 July 2020 to 17 September 2020, with a frequency of every 5 minutes. We extracted 10144 hashtags that have appeared on the HSL during this time period and traced back the original user-generated posts containing these hashtags during the time interval from birth till first appearance on the HSL. There are unavoidable problems related to the data. First, the crawling of the data was occasionally interrupted leading to loss of data. The estimated related data corruption is approximately 5\%. Censorship is another source of information loss, one example is about the pop star Wu Yifan whose account was closed and all posts related to him were no long available on Weibo ever since he was arrested by police due to several crimes~\cite{wuyifan}. Another source of data loss is due to a possibility provided by Sina Weibo, enabling users to choose privacy option, which hides their activities and makes it impossible to trace back the chain of reposts along that branch. The datasets supporting the conclusions of this article are available in the github repository, \url{https://github.com/cuihaosabrina/Emergence\_Popularity\_Sina\_Weibo}.

\subsection{Network construction} 
\begin{figure}[htbp]
\begin{center}
\includegraphics[scale=0.3]{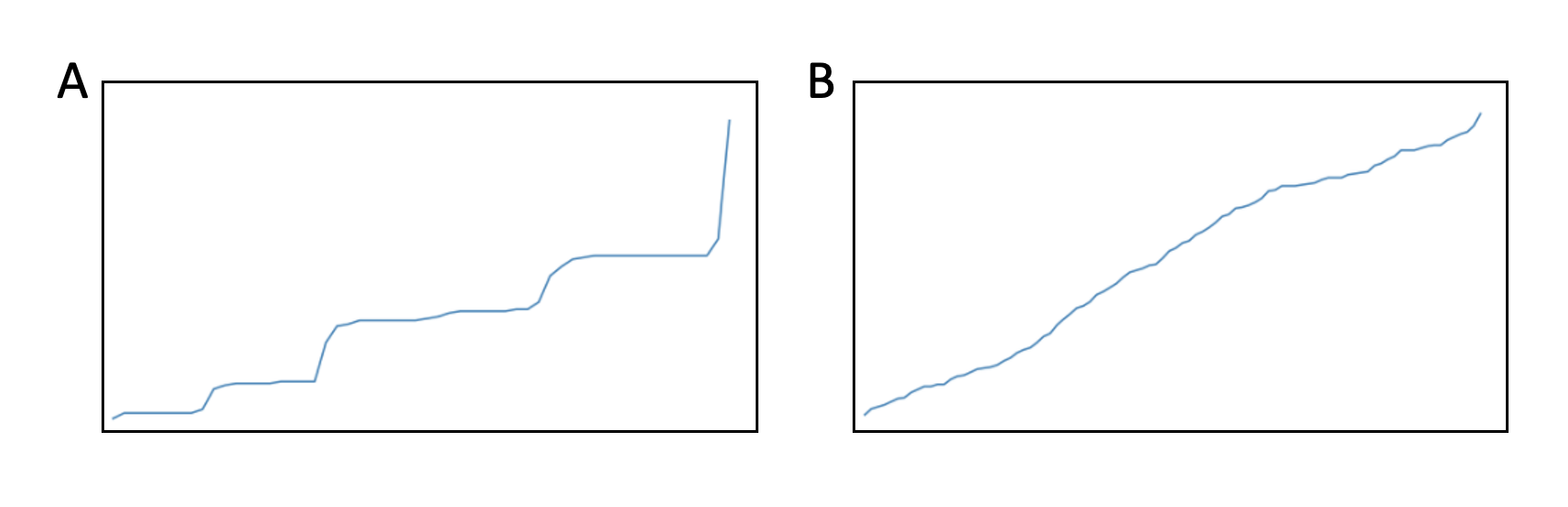}
\end{center}
\caption{Shape examples of stepwise and smooth patterns of repost network cumulative link growth trajectories.}
\label{fig:fig5}
\end{figure}

In this section, we introduce how we construct our hashtag repost networks and the properties we study. The repost network (called retweet graph in the context of Twitter) is a standard tool to study the spreading of content on microblogging sites~\cite{Thij_2014}.
The temporal directed repost network consists of users as nodes and reposts as links between them, pointing towards the user who reposts. 
{The repost networks of a hashtag contain all the origin nodes that posted this hashtag as well as the reposts of these origin nodes.}
Since all reposts have timestamps, the evolution of the repost network can be followed and can be traced back to the origins 
of the hashtag. The timestamp of the first post containing a hashtag is the birth time of that hashtag. 
We focus on the whole repost network and the largest connected component (LCC), which consists of the largest number of connected nodes in the whole hashtag repost network. When studying the network size growth, we disregard the directedness of the links. 
As the repost network evolves, the giant component identified at some stage of the evolution may be replaced by another more recently formed giant component at a later stage. In fact, this change happens often just before the hashtag almost reaches the HSL. We study the dynamics of the LCC in the repost network structure just before a hashtag reaches the HSL since it captures the most influential nodes and links during the process of the hashtag popularity emergence. For different hashtags, the growth rates of the whole network as well as the final LCC at different stages during prehistory period may have different growth characteristics as shown in Fig. \ref{fig:fig5}. We compare, analyze, and categorize these patterns.

\subsection{Classification of link growth trajectories} 

\begin{figure}[htbp]
\begin{center}
\includegraphics[scale=0.5]{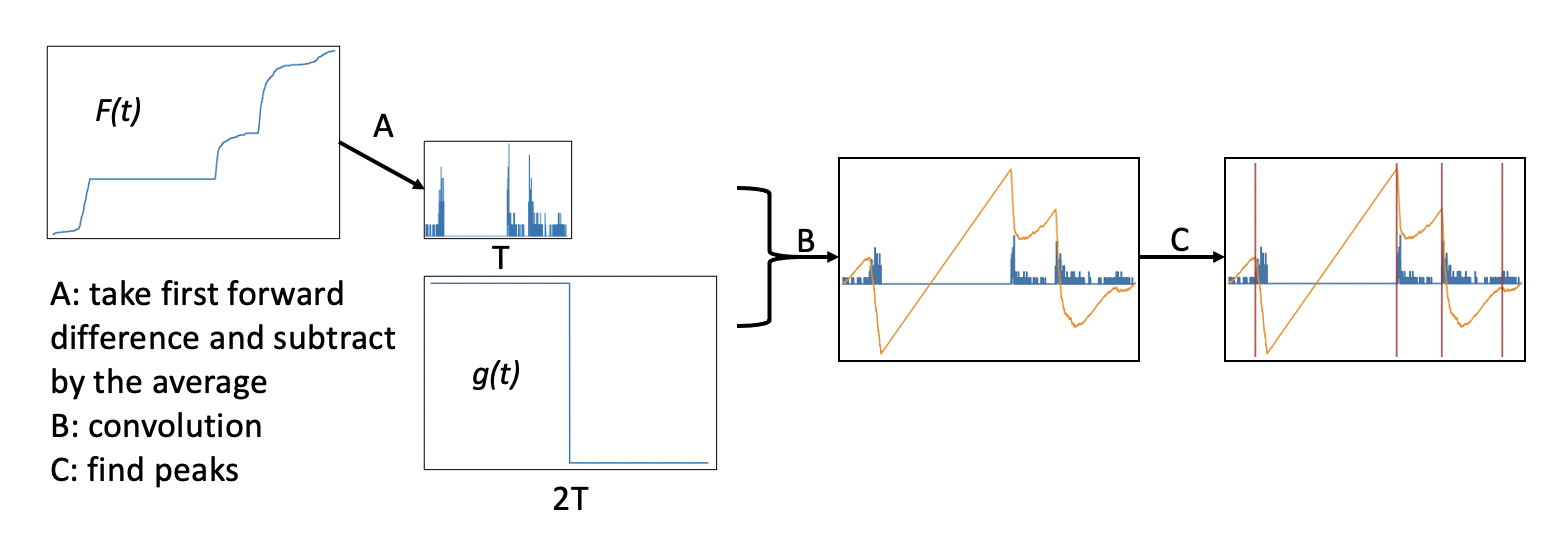}
\end{center}
\caption{Example workflow of peak detection in repost network cumulative link growth trajectory $F(t)$. $F(t)$ is a discrete function, where $t\in\mathbb{Z}, 0\le t \le T$, $T$ is the total number of minutes in the prehistory. \textbf{Step A:} $\widetilde{f} = f'(t) - \Bar{f}$, where $f'(t)$ is the first forward difference, $t\in\mathbb{N}, 1\le t \le T$, and $\Bar{f}$ is the average value of $f'(t)$. \textbf{Step B:} take the convolution $(\widetilde{f} \ast g)(t)$ where $g(t)$ is a step function. \textbf{Step C:} detect peaks by comparing with neighboring values in the convolved series.}
\label{fig:fig6}
\end{figure}



We characterize the different repost network dynamics by studying growth patterns of the cumulative number of links $F(t)$ at a minute resolution. $F(t)$ is a discrete function, where $t\in\mathbb{Z}$, $0\le t \le T$, $T$ is the total number of minutes in the prehistory. 
We use a classifier to distinguish between stepwise and smooth growth, which is based on the detection of local peaks in the derivative of the function $F(t)$. Since $F(t)$ is discrete, we obtain $f'(t)$, $t\in\mathbb{N}$, $1\le t \le T$ by taking the first forward difference of $F(t)$. Then we take the convolution of $\widetilde{f} = f'(t) - \Bar{f}$ and $g(t)$, where $\Bar{f}$ is the average value of $f'(t)$, and $g(t)$ is defined as follows 

\[ g(t) = \begin{cases} 
      1 & t\in\mathbb{N}, 1\le t \le T \\
      -1 & t\in\mathbb{N}, T+1\le t \le 2T \\
   \end{cases}
\]
In practice, we calculte the convolution $(\widetilde{f} \ast g)(t)$ using the convolve method in the numpy~\cite{numpy} Python module, with the mode parameter equals `valid'. 
We find all local maxima by comparing with neighboring values in the convolved series $(\widetilde{f} \ast g)(t)$ using the peak detection function find\_peaks in the scipy.signal~\cite{scipy} Python module. 
The principle of the classifier is demonstrated in Fig.~\ref{fig:fig6}. If there are more than two peaks identified and any of the time intervals between two consecutive peaks is greater than one hour, then we classify $F(t)$ as stepwise, otherwise smooth. The same classification procedure applies to the LCC. As for the repost network increment time series in Fig. \ref{fig:fig4}E and Fig. \ref{fig:fig4}F, the resize was done by using TimeSeriesResampler from tslearn \cite{JMLR:v21:20-091} Python package, with the method of spline interpolation~\cite{WP_spline}.



\section{Results}

We call those hashtags successful, which make it to the HSL.
The prehistory of a successful hashtag prior to entering the HSL starts from the birth of the earliest post containing this hashtag and ends at the moment this hashtag first appears on the HSL. Does the birth time of a hashtag influence the time length of its prehistory? What are the patterns of the repost network dynamics and their relation with the time length of the prehistory? To answer these questions, we summarize the observed statistical patterns of the successful hashtags that have appeared on the HSL in the observation period. 

\subsection{Role of birth time}

\begin{figure}[htbp]
\begin{center}
\includegraphics[scale=0.4]{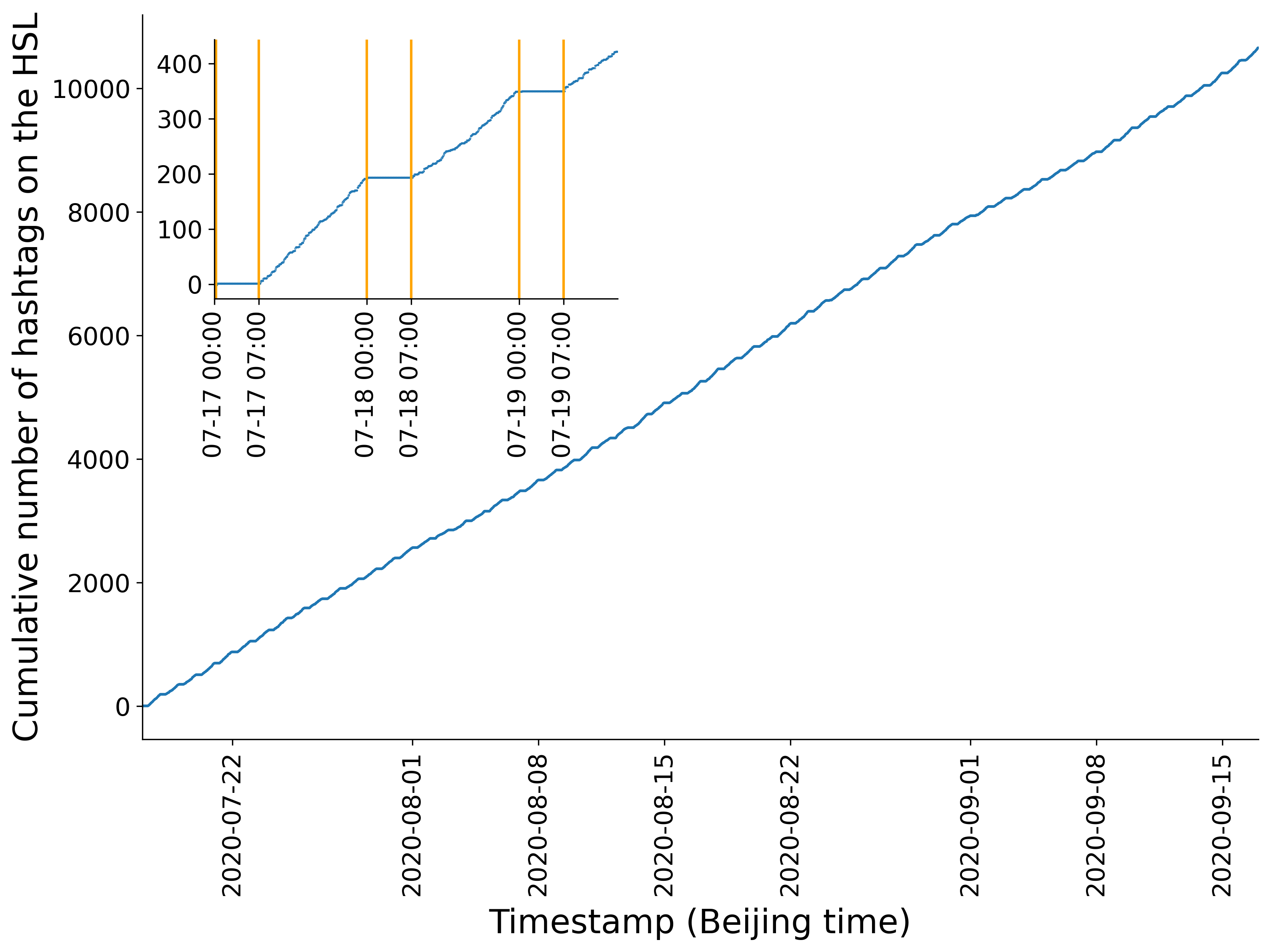}
\end{center}
\caption{Growth of cumulative number of hashtags that have ever appeared on the Sina Weibo Hot Search List (HSL) with time, from 17 July 2020 to 17 September 2020. 
{The inset enlarges the first two days and shows a clear circadian pattern as illustrated by the yellow lines. During the night time between midnight to approximately 7 am, practically no new hashtags appear on the HSL. The night time intervals are the horizontal parts in the figure, with an average size of 7.18 hours, and a standard deviation of 0.85 hours.}
}
\label{fig:fig1}
\end{figure}

{
According to its size, China should have five geographic time zones~\cite{WP_China_time} but it follows one single standard time, the Chinese (or Beijing) Standard Time. In principle, this could lead to the screening of any circadian pattern. However,
Weibo users are densely distributed in the eastern and central regions of China \cite{weibo_geography} whose geographical time zones are similar and the population accounts for 65.8 percent of the national population \cite {population_distribution}. The company Weibo Corporation has its headquarters in Beijing. In fact, we have detected clear circadian patterns.}

The cumulative number of hashtags that have ever appeared on the HSL grows approximately linearly, as shown in Fig. \ref{fig:fig1}. Zooming into the figure as seen in the inset in Fig. \ref{fig:fig1}, a periodic pattern becomes visible indicating that practically no new hashtags appear on the HSL in certain time intervals during nights.
The beginning and ending boundaries of the idle periods are sharp rather than gradual, leading to the suspicion of human control of the HSL and that the controllers' working times follow a circadian pattern. This is in contrast to the claim of Sina Weibo~\cite{weibo_announce} that the selection of hashtags to the HSL follows an automated procedure just based on a formula (see Section Methods). 

\begin{figure}[htbp]
\begin{center}
\includegraphics[scale=0.3]{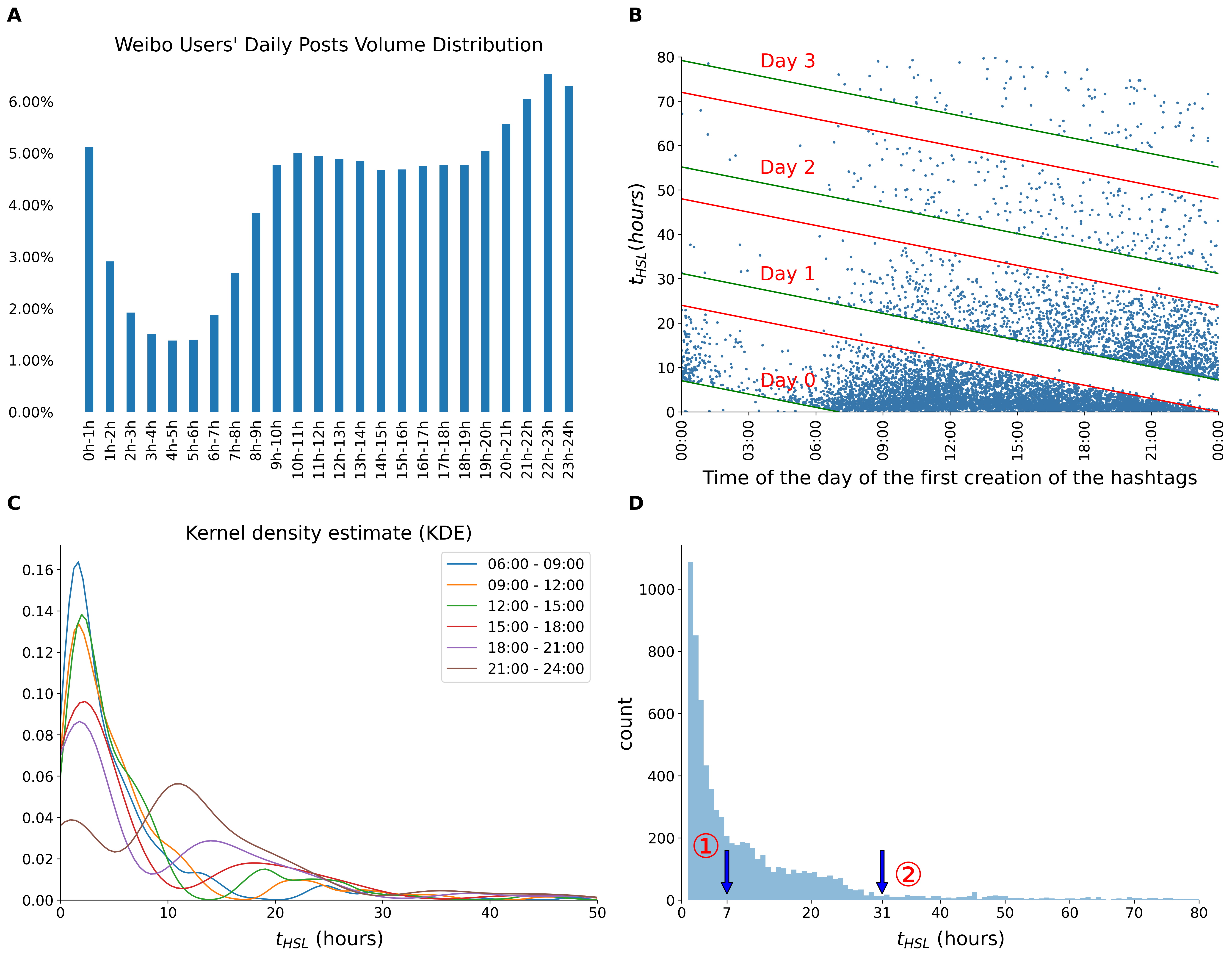}
\end{center}
\caption{Statistics of 10144 hashtags that have appeared on Sina Weibo Hot Search List (HSL) from 17 July 2020 to 17 September 2020. \textbf{(A)} Distribution of Weibo users' daily posts volume according to Weibo User Development Report~\cite{weibodata}. \textbf{(B)} Relationship between birth time of the day of the hashtags and the hours from birth to first appearance on the HSL, which we call the "HSL time" and denote as $t_{HSL}$.
The vertical difference between two lines of the same color is 24 hours, the difference of a red line and a green line on y axis is 7.18 hours. All lines are parallel. \textbf{(C)} Parameterized probability density functions of the HSL time 
by different time intervals of birth time of the day, using kernel density estimation (KDE)\cite{WP_kernel}, {with the parameter bw = "scott"\cite{scott2015multivariate}}. \textbf{(D)} Histogram of the HSL time. 
Section \textcircled{1} represents the category "Born in Rome" and section \textcircled{2} represents "Sleeping Beauty". }
\label{fig:fig2}
\end{figure}

People creating the hashtags on Weibo are largely influenced by their circadian rhythm thus the number of launched hashtags shows according variations.
Following the Weibo User Development Report~\cite{weibodata}, we show in Fig. \ref{fig:fig2}A that the number of new user-generated posts gradually increases from around 5 am, reaching the first peak around noon followed by a small decrease from 1 pm -- 2 pm, then a steady increase from 3 pm till the peak in the evening hours 10 pm -- 11 pm, and then a final decay afterwards until 5 am. 
Figure \ref{fig:fig2}B shows a scatter plot of hashtags' birth times of the day and the time length of the prehistories, which starts from the hashtag birth time until first appearance on the HSL, the ``HSL time" denoted by $t_{HSL}$. Figure \ref{fig:fig2}B shows the prehistory spanning for 4 days, separated by white stripes with vertical widths of 7.18 hours, which is the average of the night time periods in Fig. \ref{fig:fig1} when practically no new hashtags enter the HSL. A point within Day $i$ section means that the corresponding hashtag enters the HSL after $i$ days of its birth, i.e., Day 0 means it gets to the HSL within the same day as it was born. From the overall statistics in Fig. \ref{fig:fig2}B, we could see that for the hashtags whose birth time of the day is in the morning, the time it takes to enter the HSL ranges from very immediate till around 10 hours in most cases. Hashtags born from midnight to 6 am enter only exceptionally the HSL; this stripe is practically empty, indicating an idle mode with some manually introduced special cases. Figure \ref{fig:fig2}C describes the distribution of $t_{HSL}$ 
of the hashtags in Fig. \ref{fig:fig2}B parametrized on time intervals. For hashtags whose birth time of the day is after 9 pm, they will either get to the HSL in the same day within three hours, or they will show up after at least around seven hours. As the $t_{HSL}$
gets longer, the hashtags that enter the HSL become fewer. Figure \ref{fig:fig2}D shows the distribution of $t_{HSL}$, with a rapid decrease till 7 hours, followed by a slower and longer decrease afterwards.

\subsection{Roads to success} 

Successful hashtags, which make it to the HSL, may have very different prehistories. We have seen that the time of the birth of the hashtag matters as it affects the time needed to get to the HSL. In general, we observed that there are hashtags, which get very fast to the HSL and others, which need rather long time. The hashtags belonging to the first group need very short time to get to the HSL - we call this group ``Born in Rome". On the other hand, there is a group of hashtags which surpass a dormant period before discovered by a broader audience and get finally to the HSL - these are called ``Sleeping Beauty". Furthermore, we are investigating the repost network during the prehistory, and explore the differences in its evolution and topology.

\subsubsection{``Born in Rome" and ``Sleeping Beauty"}

As the proverb goes, ``All roads lead to Rome", so those already born in Rome are more likely to succeed. The name suggests that these hashtags achieve success on the HSL easily as they usually immediately gain a huge attention wave or several attention waves shortly one after the other, reaching the HSL within a few hours. The attention wave-drivers are usually superhubs or a crowd of smaller hubs.  
A superhub is an influential node whose number of followers is huge and the positioning of the account is authoritative to the type of content it posts. To name a few such superhubs, ``CCTV News"(\begin{CJK*}{UTF8}{gbsn}``央视新闻"\end{CJK*}, 126M followers), ``People's Daily"
 (\begin{CJK*}{UTF8}{gbsn}``人民日报"\end{CJK*}, 145M followers), ``Headline News"
 (\begin{CJK*}{UTF8}{gbsn}``头条新闻"\end{CJK*}, 100M followers). Successful hashtags concerning accidents, crimes, natural disasters and other societal issues (called here "social") are usually associated with the above mentioned superhubs. For hashtags related to stars and entertainment, it is more often to see the contributions of series of smaller hubs to their success. Video examples of repost network evolution in the prehistory are available in the Supplementary Information. For this type of emergence mechanism of popularity, the time for a successful hashtag to enter the HSL is usually short. As Fig. \ref{fig:fig1} and Fig. \ref{fig:fig2}B suggests, the closer the hashtags are born to midnight, the less likely they tend to appear on the HSL immediately, instead, they tend to show up after at least seven hours, making their prehistory longer. To factor out the influence of night hours, we consider hashtags whose time needed from birth till HSL within 7 hours to be in the category "Born in Rome". As shown in Fig. \ref{fig:fig2}D, at 7 hours we see the start of a shoulder which marks change in the shape of the count of hashtags versus waiting time. We have also seen some hashtags with very few (re)posts prior to the HSL, for example,  \#US orders 100 million doses of coronavirus vaccine from UK and France\# (\begin{CJK*}{UTF8}{gbsn}\#美国从英法订购1亿剂新冠疫苗\#\end{CJK*}), which could result from human intervention regarding international news. 

We call another type of successful hashtags ``Sleeping Beauty", when the emergence mechanism results in relatively long time needed from birth till HSL. Hashtags in this category usually experience a low activity dormant period before being picked up by crucial influencing nodes. They might need several attention waves, and that the inter-wave time intervals can be long before a final trigger of significant popularity pushing them to the HSL. As marked in Fig. \ref{fig:fig2}D, at around 31 hours the count of hashtags drops to a very low-level plateau. We use 31 hours (one day plus seven hours inactive night period) as the boundary for ``Sleeping Beauty", indicating that the delay is substantial and not due to the night break. When it comes to the hashtag content, as shown in Fig. \ref{fig:fig3}, ``Sleeping Beauty" exhibits a higher proportion of the Stars and lower proportion of Social and International categories than ``Born in Rome". See classification details in the Supplementary Information.


For the ``Sleeping Beauty" category, as $t_{HSL}$ increases, it is more likely to experience the ``rebirth" of the same hashtag, so that the hashtags generated at a later time might not refer to the same event at the birth of the hashtag, though the hashtag itself remains unchanged. The examples are shown in the Supplementary Information. In order to avoid such cases, we restricted the ``Sleeping Beauty" category to those with $t_{HSL}<5$ days, resulting in altogether 571 hashtags in this category and crawled all their reposts. In addition, we produced an equal-sized random sample from the ``Born in Rome" category. We crawled the number of followers of all users who participated in the posting behavior, with 69k users in total.

\subsubsection{Relation with repost network dynamics} 
\begin{figure}[htbp]
\begin{center}
\includegraphics[scale=0.5]{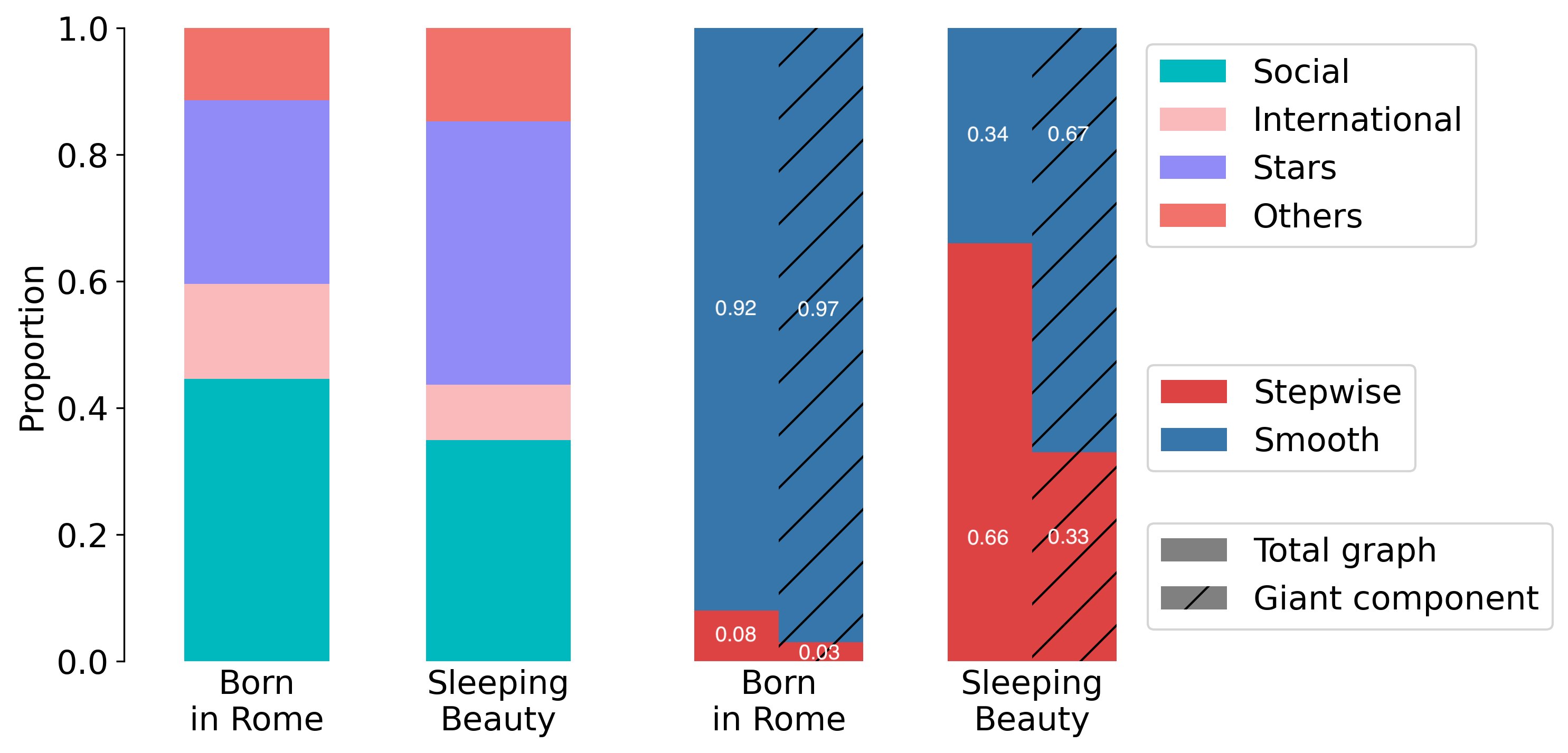}
\end{center}
\caption{Distribution of hashtags from ``Sleeping Beauty" category and the same size random sample from ``Born in Rome" category by hashtag content and the shape of their link growth patterns, whether stepwise or smooth, of the whole repost network as well as the final giant component.}
\label{fig:fig3}
\end{figure}

\begin{figure}[htbp]
\begin{center}
\includegraphics[scale=0.35]{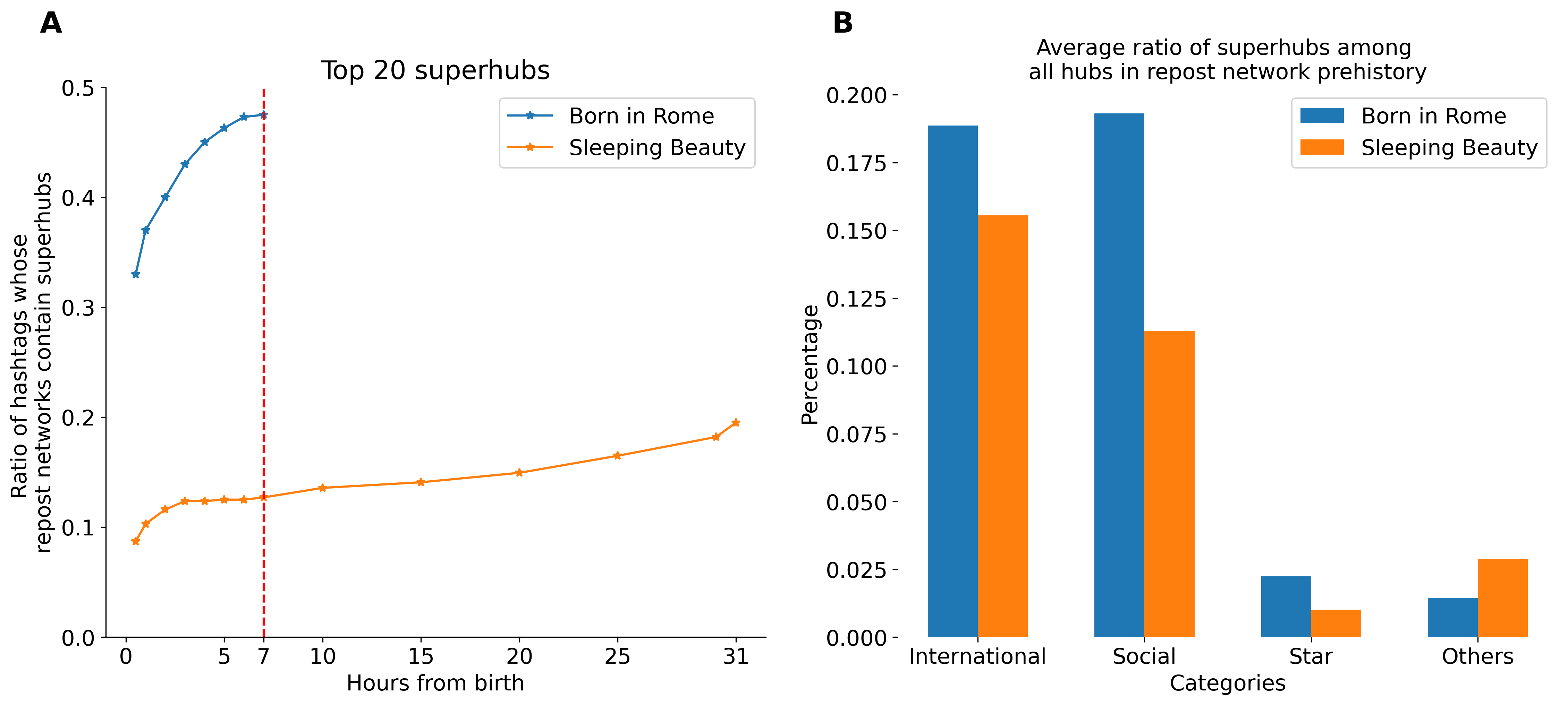}
\end{center}
\caption{Role of superhubs for hashtags in ``Born in Rome" and ``Sleeping Beauty" categories. \textbf{(A)} Ratio of hashtags whose repost networks in the prehistory contain at least one superhub (top 20 largest hubs excluding celebrities).  \textbf{(B)} Ratio of superhubs among all hubs (nodes with top 10k largest degrees) in the prehistory. 
}
\label{fig:fig44}
\end{figure}


The hashtag repost network grows in time as we define it as the cumulative (or aggregate) network of the reposts of online users. Different repost networks vary in growth speeds and topological structures. We have studied the repost network dynamics of hashtags in the ``Born in Rome" and ``Sleeping Beauty" categories. Fig. \ref{fig:fig3} shows the ratio of different link growth pattern dynamics of the total network and the final giant component in the two categories (for examples see Fig. \ref{fig:fig5}). As shown in Fig. \ref{fig:fig3}, for the total repost network growth, the majority of ``Sleeping beauty" have stepwise shape, meaning the necessity of several attention waves to gain the popularity to enter the HSL. As for the ``Born in Rome" category, the majority hashtags have smooth shape in the repost network link growth, meaning that the power of the hub(s) at their early stage is enough to push the hashtags to reach system-wide popularity. The ratio of hashtags influenced by superhubs (top 20 largest hubs excluding celebrities, see SI) in the prehistory is a function of time measured from the birth. As shown in Fig. \ref{fig:fig44}A, this ratio starts at a higher value for ``Born in Rome" category and increases rapidly, while for ``Sleeping Beauty" category, it remains at a relatively low level during the whole prehistory period. As shown in Fig. \ref{fig:fig44}B, superhubs play a more important role in the categories of International and Social, while for the Star category, smaller hubs are dominant. The proportions of stepwise shape in the giant components of both categories are fewer than those of the total graph. This is reasonable since the formation starting time of the final giant component could be later than that of the whole repost network.

\subsubsection*{Failure and success} 

\begin{figure}[htbp]
\begin{center}
\includegraphics[scale=0.35]{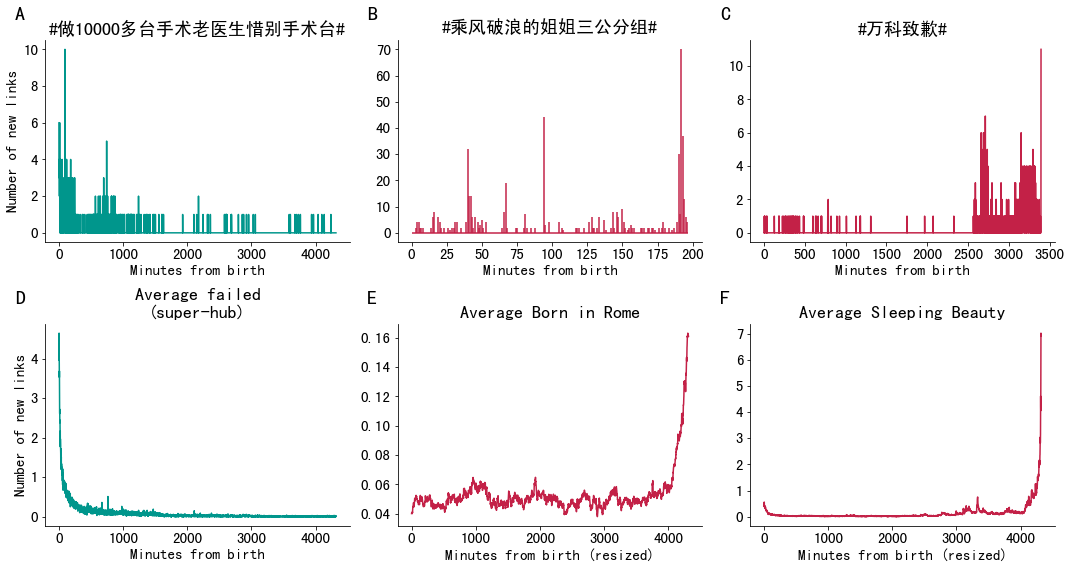}
\end{center}
\caption{Comparison of repost network growth patterns of failed hashtags born in the superhub "CCTV News" (\begin{CJK*}{UTF8}{gbsn}``央视新闻" \end{CJK*}) as well as successful hashtags from the categories ``Born in Rome" and ``Sleeping Beauty". Note the different time scales in the figures. \textbf{(A)} An example of a failed hashtag born in the superhub ``CCTV News", \#Doing more than 10000 operations, the old doctor bid farewell to the operating table\# (\begin{CJK*}{UTF8}{gbsn}\#做10000多台手术老医生惜别手术台\#\end{CJK*}). \textbf{(B)} An example of a hashtag from the category ``Born in Rome", \#Sisters Who Make Waves (a variety show) grouping of the third public performance\# (\begin{CJK*}{UTF8}{gbsn}\#乘风破浪的姐姐三公分组\#\end{CJK*}). \textbf{(C)} An example of a hashtag from the category ``Sleeping Beauty", ,\#Vanke apologizes\# (\begin{CJK*}{UTF8}{gbsn}\#万科致歉\#\end{CJK*}).
\textbf{(D)} Average repost network growth pattern of 100 randomly selected failed hashtags from the superhub ``CCTV News" in late August 2020, lasting for three days (4320 minutes) from birth time. 
\textbf{(E)} Average repost network growth pattern of all ``Born in Rome" sample hashtags, time length resized to 4320 for all hashtags.
\textbf{(F)} Average repost network growth pattern of all ``Sleeping Beauty" hashtags, time length resized to 4320 for all hashtags.} 
\label{fig:fig4}
\end{figure}

Hubs or superhubs are needed for a hashtag to reach popularity, however, not all hashtags born in superhubs are successful as many  - in fact, the majority - of them fail to land on the HSL. How does the growth pattern of the repost network of unsuccessful hashtags differ from those of successful ones? We took the superhub \#CCTV News\# as an example and studied the repost network evolution of 100 randomly selected hashtags in late August 2020. One example is shown in Fig. \ref{fig:fig4}A, the hashtag first attracted considerable attention, and then the attention decreased in a fluctuating manner and the temporary gains were not enough to compete with other hashtags for a position on the HSL. The averaged repost network growth pattern of the unsuccessful hashtags born in \#CCTV News\# is shown in Fig. \ref{fig:fig4}D, in minute resolution. The network increment per minute shows a fast (exponential) decay and then a slower one as time goes on. In Fig. \ref{fig:fig4}B and Fig. \ref{fig:fig4}C, we show examples of hashtags from ``Born in Rome" and ``Sleeping Beauty" categories respectively. One or several attention waves are launched before the hashtags reach HSL, and the number of new links generally shows an increasing trend. Figure \ref{fig:fig4}E and Fig. \ref{fig:fig4}F show the averaged repost network growth patterns of ``Born in Rome" and ``Sleeping Beauty" hashtags respectively, with the time length resized to three days. The fast decay in the early time behavior of the averaged ``Sleeping Beauty" curve is very similar to that of the unsuccessful ones, as shown in Fig \ref{fig:fig4}D. The higher initial value for the unsuccessful hashtags is due to the fact that we selected the unsuccessful hashtags from those starting at the superhub ``CCTV News" which assured considerable early attention, while for the ``Sleeping Beauty" hashtags we took all cases, irrespective of the popularity of the node where the hashtags were born. In contrast to the unsuccessful hashtags, ``Sleeping Beauties" experience at a later stage a push in the attention dynamics due to getting picked up by a large hub which finally help them to get to the HSL.

\section{Discussion}

In this paper, we examined the emergence of hashtag popularity on the Chinese microblogging site Sina Weibo by analyzing prehistory of the repost network evolution of hashtags that finally get to the HSL. We have focused on the HSL time $t_{HSL}$ and studied differences in the repost network dynamics of the whole network as well as of the final giant component for successful hashtags. Our studies have identified two extreme types of popularity emergence mechanisms for successful hashtags: That of the ``Born in Rome" and of the ``Sleeping Beauty", and pointed out the role of different hubs in the process. Compared with ``Sleeping Beauty" hashtags, those in ``Born in Rome" category tend to reach superhubs at an early stage of their spreading process, facilitating their success to the HSL. For ``Sleeping Beauty" hashtags, instead of reaching superhubs at an early stage, they are likely to gain several attention waves from smaller hubs, resulting often in a stepwise growth pattern of the repost network.

Previous studies on the emergence of popularity of entities in online text streams observed two patterns: a ``bursty" one where content blasts into activity in public discourse without a precedent, and a ``delayed" pattern which experiences a period of inactivity before resurfacing \cite{graus2018birth}. These patterns are similar to the ``Born in Rome" and ``Sleeping Beauty" discussed in this paper. As Fig. \ref{fig:fig3} shows, two thirds of sleeping beauty hashtags exhibit a stepwise shape of repost network growth, meaning that there are at least two peaks before reaching the HSL and there is considerable time delay between the peaks. In the case of two-peaks, the hashtag first experiences a hibernation period with low activity after birth, then a peak, then another hibernation, then a final peak and gets finally to the HSL. Our findings about sleeping beauties is in general in accordance with previous studies~\cite{zhang2017sleeping}, though we consider only the prehistory phase of information diffusion. We also notice that in Fig. \ref{fig:fig3} around one third of sleeping beauty hashtags have smooth shape, meaning they only experience one long hibernation period before a peak that propel them to the HSL. For the final giant component growth pattern, the majority of hashtags have smooth shape, they experience one hibernation period before a final peak to reach HSL, implying the time period closer to the appearance on HSL is more important in determining the temporal popularity that makes the hashtags enter the HSL. For sleeping beauties, strategically locating or intervening the tipping point(s) could contribute to marketing efficiency and future popularity prediction~\cite{zhang2017sleeping}, or to destroy the formation of trends in the case of misinformation.

When it comes to the content categories of the hashtags, we observe differences between Star and Social/International. For both ``Born in Rome" and ``Sleeping Beauty" hashtags, International and Social hashtags tend to have higher ratios of superhubs than the Star hashtags. For the Star hashtags, non-superhubs play a more important role in their popularity. In fact, it is often the case that the repost network evolution of hashtags related, e.g., to celebrities results from ``collective efforts" and their popularity accounts for concerted influence of several smaller hubs which are usually marketing accounts. This emphasizes the importance of social capital in making hashtags related to stars popular enough to enter the HSL.

Though superhubs are important in triggering hashtag popularity, by far not all hashtags created by the most prominent superhubs make it to the HSL. The timing of the first creation of a hashtag is an important factor to its popularity evolution, since it influences the system-wide user attention level as well as the pool of the competing hashtags.
From the statistics shown in Fig. \ref{fig:fig2}A, the volume of user posts from midnight to 6 am is at most a factor of 2 lower than during a similar period in the rest of the day, while the proportion of successful hashtags in the same time period is significantly less as Fig. \ref{fig:fig2}B suggests, indicating the disadvantage of hashtags born in that time period in achieving success. 
Sina Weibo Hot Search List is of commercial value with an advertising effect on the hashtags by substantially increasing their visibility to the whole public.
Understanding the mechanism of the emergence of hashtag popularity and the importance of timing, on the one hand, could contribute to marketing and maximizing the spreading efficiency by playing with these factors. On the other hand, it provides Weibo users with better knowledge to differentiate about the possible social capital influence in promoting certain contents, such as Star hashtags.

It is of great social value and importance to have hashtags on HSL that raise real public awareness and concern. Of course, even hashtags from superhubs could fail to get to the list, let alone those from regular users with only a few followers. For hashtags in the latter category, it is hard to be successful. In fact, hashtags posted by normal users need to be (re)posted by influential ones to be promoted enough, leading to the necessity that influential nodes should get aware of their social responsibilities to participate in such situations where the voice of unprivileged people are unheard by the whole audience. More importantly, the prime responsibility is carried by the platform provider. It is challenging but important to use a fair algorithm to take into account of the opinion of the ``invisible majority" (in terms of number of followers) and capture real hot topics that gain true attention from the public. In reality, there are signs that - in spite of the claims by Sina Weibo - the selection of the hashtags to the list is not entirely automated. One clear signature of this is the night break during which practically no new hashtags appear on HSL, however, time to time some do.

Despite that this study focuses on the emergence mechanism of the hashtag popularity specific to Weibo, we mention that our approach may shed some lights on more than just for this online microblogging complex system. Generally, for any successful cultural product, such as a song, a TV series, a best-seller book, etc., there is also a prehistory prior to the success when attention of the broad society is reached. During that prehistory, people interact with each other in relation to this product, for example, recommend, comment and consume. Such processes are in the focus of the study of innovations~\cite{Rogers2010Innovations}.
What interaction mechanisms lead to the success of a cultural product? Does the birth timing of this cultural product influence the time length it takes to achieve success? What are the differences in the popularity mechanisms between products from ``Born in Rome" and ``Sleeping Beauty" types if there are any? What are the components when forming the attention waves if there are several? Of course, the time scale for such products is very different from that of the hashtags, but the online digital ``footprints" of the related social interactions could be very helpful in uncovering the important details of the processes. 





\section*{Acknowledgements} 
We acknowledge support supported by the European Union – Horizon 2020 Program under the scheme ``INFRAIA-01-2018-2019 – Integrating Activities for Advanced Communities", Grant Agreement n.871042, ``SoBigData++: European Integrated Infrastructure for Social Mining and Big Data Analytics" and
SAI enabled by FWF (I 5205-N) within the EU CHIST-ERA program.

\bibliographystyle{elsarticle-num} 
\bibliography{sample.bib}

\begin{thebibliography}{10}
\expandafter\ifx\csname url\endcsname\relax
  \def\url#1{\texttt{#1}}\fi
\expandafter\ifx\csname urlprefix\endcsname\relax\def\urlprefix{URL }\fi
\expandafter\ifx\csname href\endcsname\relax
  \def\href#1#2{#2} \def\path#1{#1}\fi

\bibitem{zhang2016creates}
L.~Zhang, J.~Zhao, K.~Xu, Who creates trends in online social media: The crowd
  or opinion leaders?, Journal of Computer-Mediated Communication 21~(1) (2016)
  1--16.

\bibitem{bao2013popularity}
P.~Bao, H.-W. Shen, J.~Huang, X.-Q. Cheng, Popularity prediction in
  microblogging network: a case study on sina weibo, in: Proceedings of the
  22nd international conference on world wide web, 2013, pp. 177--178.

\bibitem{ma2013towards}
H.~Ma, W.~Qian, F.~Xia, X.~He, J.~Xu, A.~Zhou, Towards modeling popularity of
  microblogs, Frontiers of Computer Science 7~(2) (2013) 171--184.

\bibitem{goel2016structural}
S.~Goel, A.~Anderson, J.~Hofman, D.~J. Watts, The structural virality of online
  diffusion, Management Science 62~(1) (2016) 180--196.

\bibitem{Annamoradnejad2019twitter_trend}
I.~Annamoradnejad, J.~Habibi, A comprehensive analysis of twitter trending
  topics, in: International Conference on Web Research (ICWR), 2019, pp.
  22--27.

\bibitem{ch_jk}
H.~Cui, J.~Kert{\'e}sz, Attention dynamics on the chinese social media sina
  weibo during the covid-19 pandemic, EPJ data science 10~(1) (2021) 8.

\bibitem{asur_2011}
S.~Asur, B.~A. Huberman, G.~Szabo, C.~Wang, Trends in social media: Persistence
  and decay, in: Proceedings of the International AAAI Conference on Web and
  Social Media, Vol.~5, 2011, pp. 434--437.

\bibitem{Thij_2014}
M.~t. Thij, T.~Ouboter, D.~Worm, N.~Litvak, H.~v.~d. Berg, S.~Bhulai, Modelling
  of trends in twitter using retweet graph dynamics, in: International Workshop
  on Algorithms and Models for the Web-Graph, Springer, 2014, pp. 132--147.

\bibitem{Ratkiewicz_2011}
J.~Ratkiewicz, M.~Conover, M.~Meiss, B.~Gon{\c{c}}alves, S.~Patil, A.~Flammini,
  F.~Menczer, Truthy: mapping the spread of astroturf in microblog streams, in:
  Proceedings of the 20th international conference companion on World wide web,
  2011, pp. 249--252.

\bibitem{twitter}
Q4 and fiscal year 2021 letter to shareholders, \emph{Twitter}
  \url{https://s22.q4cdn.com/826641620/files/doc_financials/2021/q4/Final-Q4'21-Shareholder-letter.pdf}
  (2022).

\bibitem{mau}
Number of monthly active users of sina weibo from 1st quarter of 2018 to 3rd
  quarter of 2021, \emph{statista}
  \url{https://www.statista.com/statistics/795303/china-mau-of-sina-weibo/}
  (2021).

\bibitem{jiang_fan}
R.~Staff, China punishes microblog platform weibo for interfering with
  communication, \emph{Reuters}
  \url{https://www.reuters.com/article/us-china-censorship-weibo-idUSKBN23H1J2}
  (2020).

\bibitem{censor}
L.~Chen, C.~Zhang, C.~Wilson, Tweeting under pressure: analyzing trending
  topics and evolving word choice on sina weibo, in: Proceedings of the first
  ACM conference on Online social networks, 2013, pp. 89--100.

\bibitem{censor2}
J.~A. Vuori, L.~Paltemaa, The lexicon of fear: Chinese internet control
  practice in sina weibo microblog censorship, Surveillance \& society 13~(3/4)
  (2015) 400--421.

\bibitem{Ma_2013}
H.~Ma, W.~Qian, F.~Xia, X.~He, J.~Xu, A.~Zhou, Towards modeling popularity of
  microblogs, Frontiers of Computer Science 7~(2) (2013) 171--184.

\bibitem{Romero_2011}
D.~M. Romero, B.~Meeder, J.~Kleinberg, Differences in the mechanics of
  information diffusion across topics: idioms, political hashtags, and complex
  contagion on twitter, in: Proceedings of the 20th international conference on
  World wide web, 2011, pp. 695--704.

\bibitem{Tsur_2012}
O.~Tsur, A.~Rappoport, What's in a hashtag? content based prediction of the
  spread of ideas in microblogging communities, in: Proceedings of the fifth
  ACM international conference on Web search and data mining, 2012, pp.
  643--652.

\bibitem{Lehmann_2012}
J.~Lehmann, B.~Gon{\c{c}}alves, J.~J. Ramasco, C.~Cattuto, Dynamical classes of
  collective attention in twitter, in: Proceedings of the 21st international
  conference on World Wide Web, 2012, pp. 251--260.

\bibitem{Pervin_2015}
N.~Pervin, T.~Q. Phan, A.~Datta, H.~Takeda, F.~Toriumi, Hashtag popularity on
  twitter: Analyzing co-occurrence of multiple hashtags, in: International
  Conference on Social Computing and Social Media, Springer, 2015, pp.
  169--182.

\bibitem{Ma_2013_prediction}
Z.~Ma, A.~Sun, G.~Cong, On predicting the popularity of newly emerging hashtags
  in t witter, Journal of the American Society for Information Science and
  Technology 64~(7) (2013) 1399--1410.

\bibitem{Yu_2020}
H.~Yu, Y.~Hu, P.~Shi, A prediction method of peak time popularity based on
  twitter hashtags, IEEE Access 8 (2020) 61453--61461.

\bibitem{Khan_2021}
H.~U. Khan, S.~Nasir, K.~Nasim, D.~Shabbir, A.~Mahmood, Twitter trends: A
  ranking algorithm analysis on real time data, Expert Systems with
  Applications 164 (2021) 113990.

\bibitem{yu2012artificial}
L.~Yu, S.~Asur, B.~A. Huberman, Artificial inflation: The true story of trends
  in sina weibo, arXiv preprint arXiv:1202.0327 (2012).

\bibitem{wu2022revealing}
L.~Wu, J.~Qi, N.~Shi, J.~Li, Q.~Yan, Revealing the relationship of topics
  popularity and bursty human activity patterns in social temporal networks,
  Physica A: Statistical Mechanics and its Applications 588 (2022) 126568.

\bibitem{zhang2017sleeping}
L.~Zhang, K.~Xu, J.~Zhao, Sleeping beauties in meme diffusion, Scientometrics
  112~(1) (2017) 383--402.

\bibitem{Zahng:2017}
Y.~Zhang, Microblogging and its implications to chinese civil society and the
  urban public sphere: A case study of sina weibo, PhD Thesis at the University
  of Queensland (2016).

\bibitem{BBC}
D.~Hewitt., Weibo brings changes to china, \emph{BBC News}
  \url{https://www.bbc.com/news/magazine-18773111} (2012).

\bibitem{weibo_announce}
W.~Administrator, Weibo hot search regulation rules, \emph{Sina Weibo}
  \url{https://weibo.com/1934183965/KuKyPkp8Y?type=repost} (2021).

\bibitem{wuyifan}
Kris wu sex scandal, \emph{Wikipedia}
  \url{https://en.wikipedia.org/wiki/Kris\_Wu\_sex\_scandal} (2021).

\bibitem{numpy}
C.~R. Harris, K.~J. Millman, S.~J. van~der Walt, R.~Gommers, P.~Virtanen,
  D.~Cournapeau, E.~Wieser, J.~Taylor, S.~Berg, N.~J. Smith, R.~Kern, M.~Picus,
  S.~Hoyer, M.~H. van Kerkwijk, M.~Brett, A.~Haldane, J.~F. del R{\'{i}}o,
  M.~Wiebe, P.~Peterson, P.~G{\'{e}}rard-Marchant, K.~Sheppard, T.~Reddy,
  W.~Weckesser, H.~Abbasi, C.~Gohlke, T.~E. Oliphant,
  \href{https://doi.org/10.1038/s41586-020-2649-2}{Array programming with
  {NumPy}}, Nature 585~(7825) (2020) 357--362.
\newblock \href {https://doi.org/10.1038/s41586-020-2649-2}
  {\path{doi:10.1038/s41586-020-2649-2}}.
\newline\urlprefix\url{https://doi.org/10.1038/s41586-020-2649-2}

\bibitem{scipy}
P.~Virtanen, R.~Gommers, T.~E. Oliphant, M.~Haberland, T.~Reddy, D.~Cournapeau,
  E.~Burovski, P.~Peterson, W.~Weckesser, J.~Bright, S.~J. {van der Walt},
  M.~Brett, J.~Wilson, K.~J. Millman, N.~Mayorov, A.~R.~J. Nelson, E.~Jones,
  R.~Kern, E.~Larson, C.~J. Carey, {\.I}.~Polat, Y.~Feng, E.~W. Moore,
  J.~{VanderPlas}, D.~Laxalde, J.~Perktold, R.~Cimrman, I.~Henriksen, E.~A.
  Quintero, C.~R. Harris, A.~M. Archibald, A.~H. Ribeiro, F.~Pedregosa, P.~{van
  Mulbregt}, {SciPy 1.0 Contributors}, {{SciPy} 1.0: Fundamental Algorithms for
  Scientific Computing in Python}, Nature Methods 17 (2020) 261--272.
\newblock \href {https://doi.org/10.1038/s41592-019-0686-2}
  {\path{doi:10.1038/s41592-019-0686-2}}.

\bibitem{JMLR:v21:20-091}
R.~Tavenard, J.~Faouzi, G.~Vandewiele, F.~Divo, G.~Androz, C.~Holtz, M.~Payne,
  R.~Yurchak, M.~Ru{\ss}wurm, K.~Kolar, E.~Woods,
  \href{http://jmlr.org/papers/v21/20-091.html}{Tslearn, a machine learning
  toolkit for time series data}, Journal of Machine Learning Research 21~(118)
  (2020) 1--6.
\newline\urlprefix\url{http://jmlr.org/papers/v21/20-091.html}

\bibitem{WP_spline}
Spline interpolation, \emph{Wikipedia}
  \url{https://en.wikipedia.org/wiki/Spline_interpolation} (2022).

\bibitem{WP_China_time}
Time in china, \emph{Wikipedia}
  \url{https://en.wikipedia.org/wiki/Time_in_China} (2022).

\bibitem{weibo_geography}
S.~W.~D. Center, Weibo 2020 user development report., \emph{Weibo Report.}
  \url{https://data.weibo.com/report/reportDetail?id=456} (2021).

\bibitem{population_distribution}
Main data of the seventh national population census, \emph{National Bureau of
  Statistics of China}
  \url{http://www.stats.gov.cn/english/PressRelease/202105/t20210510_1817185.html}
  (2021).

\bibitem{weibodata}
S.~W.~D. Center, 2015 weibo user development report., \emph{Weibo Report.}
  \url{https://data.weibo.com/report/reportDetail?id=333} (2016).

\bibitem{WP_kernel}
Kernel density estimation, \emph{Wikipedia}
  \url{https://en.wikipedia.org/wiki/Kernel_density_estimation} (2022).

\bibitem{scott2015multivariate}
D.~W. Scott, Multivariate density estimation: theory, practice, and
  visualization, John Wiley \& Sons, 2015.

\bibitem{graus2018birth}
D.~Graus, D.~Odijk, M.~de~Rijke, The birth of collective memories: Analyzing
  emerging entities in text streams, Journal of the Association for Information
  Science and Technology 69~(6) (2018) 773--786.

\bibitem{Rogers2010Innovations}
E.~M. Rogers, Diffusion of Innovations, The Free Press, New York, 2010.

\end{thebibliography}




\section*{Supplementary material}

# Supplementary Information

# "Born in Rome" or "Sleeping Beauty": Emergence of hashtag popularity on the Chinese microblog Sina Weibo

Hao Cui[1] and János Kertész[1*]

[1]Department of Network and Data Science, Central European University, Quellenstrasse 51, A-1100 Vienna, Austria

*Correspondence: kerteszj@ceu.edu

## SI1. Evolution of repost networks before getting to the HSL (movies)

Here we show how the repost networks of different categories of hashtags evolve before they get to the HSL [1].

## SI2. Hashtag categorization

We have classified the hashtags into the following categories based on human judgement: Social, Stars, International, and Others. The Social category consists of hashtags that are related to social accidents, crimes, natural disasters, and other events that are related to social life. The Stars category consists of movie/sports stars, singers, idols, celebrities as well as the TV programs/movies and events that they participate in. The International category consists of hashtags whose content is related to news outside of China. The Others category consists of the rest of the hashtags that fall into none of the above categories. The following table contains example hashtags and their translation.

| Social | Stars | International | Others |
|---|---|---|---|
| #Pensions in 31 provinces have all risen# (#31省份养老金已全部上涨#) | #Liang Zhengxian Master of Space Management# (#梁正贤空间管理大师#) | #Employees in White House cafes diagnosed with covid# (#白宫内部咖啡厅员工确诊新冠#) | #If all creatures became cats# (#假如所有生物都变成猫#) |
| #76-year-old man rescues 200-pound drowning man# | #Song Dandan's 60th birthday party# (#宋丹丹60岁生日宴 | #UAE to fully normalize relations with Israel# | #My Youth Pain# (#我的青春疼痛#) |

| (#76岁老人救起200多斤溺水者#) | #) | (#阿联酋将与以色列实现关系全面正常化#) | |
| --- | --- | --- | --- |
| #Shanxi police cracked down on intentional homicide 30 years ago# (#山西警方破获30年前故意杀人案#) | #Jay Chou's old photos from 20 years ago# (#周杰伦晒20年前旧照#) | #Melbourne, Australia will implement curfew# (#澳大利亚墨尔本将实施宵禁#) | #How to gracefully carry a quilt to school# (#如何优雅地背着被子去学校#) |

**Supplementary Table 1.** Examples of content categorization.

## SI3. Hashtag rebirth

As the prehistory length $t_{HSL}$ increases, it is more likely to experience the hashtag "rebirth", that the hashtag posted after a few days might not refer to the same event as the birth of the hashtag, though the hashtag itself remains unchanged. We show some examples here.

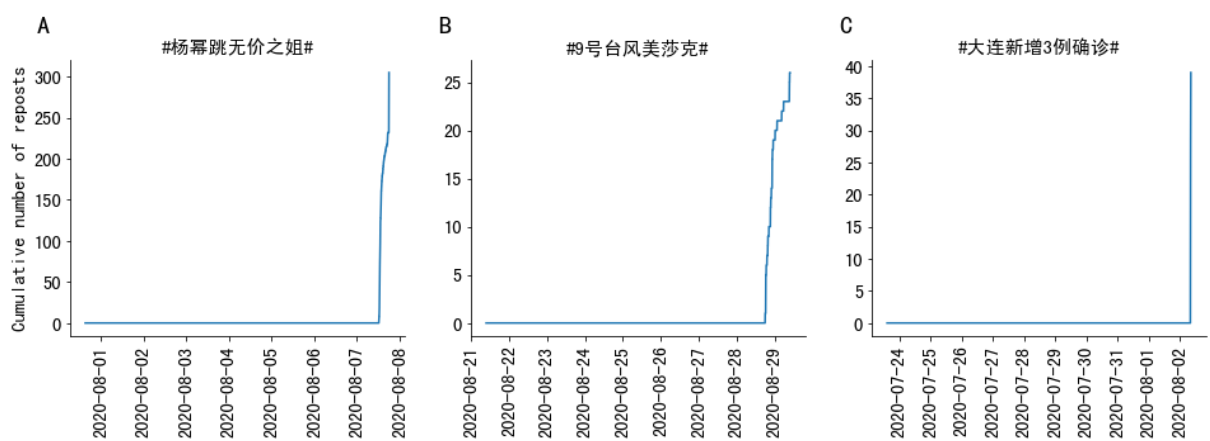

**Supplementary figure S1.** Examples of hashtag "rebirth".

- As shown in supplementary figure S1A, the hashtag #Yang Mi dances Priceless Sister# (#杨幂跳无价之姐#) first appeared on the HSL on 2020 August 7 at 17:57. It was born on 2020 July 31 15:24, with the content about the trailer of a TV show that the celebrity Yang Mi would dance Priceless Sister in the next episode, no reposts. The second post containing this hashtag was posted on 2020 August 7 12:09 when the TV show actually started. The hashtag at the birth and the hashtag created on August 7 refer to different

sources. The success of the hashtag on the HSL was the result of the TV show on August 7 instead of the trailer one week ago.

- As shown in supplementary figure S1B, the hashtag #Typhoon Maysak No. 9 # (#9 号台风美莎克#) first appeared on the HSL on 2020 August 29 at 09:54. It was born on August 21 at 09:25, with the content about the possibility that Typhoon Maysak might be coming soon. From 2020 August 21 15:42 on, no new (re)posts until 2020 August 28 when the Typhoon really formed. The posts containing the same hashtag at the later time refers to the real Typhoon rather than the warning at the beginning.

- As shown in supplementary figure S1C, the hashtag #3 new infected cases confirmed in Dalian# (#大连新增3例确诊#) first appeared on the HSL on 2020 August 2 at 08:17. It was born on 2020 July 23 14:43. Though the hashtag remains the same, the content at the birth refers to the new infected cases at that time and the hashtag on 2020 August 2 refers to the new infected cases on August 1.

As the prehistory gets longer, it is more common to see the "rebirth" of the same hashtags referring to a different event from birth. In order to avoid such influences, when studying the properties of hashtags in the "Sleeping Beauty" category, we choose those hashtags whose $t_{HSL}$ < 5 days.

## SI3. Division of Rome and Beauty

We choose hashtags whose $t_{HSL}$ < 7 hours to be in the category of "Born in Rome", and 31 hours < $t_{HSL}$ < 5 days to be in the category of "Sleeping Beauty", with the following reasons:

(1) As shown in Fig. 1, and Fig. 2B, there is a period of around 7 hours during the night when almost no new hashtags appear on the HSL. If a hashtag was born close to midnight, then it is very likely that it will experience the 7-hour break and automatically not be in the category of "Born in Rome". Similarly, for "Sleeping Beauty", we use 24+7 = 31 as the boundary, indicating that the delay is substantial and not due to the night break.
(2) The limiting points 7 and 31 hours are shown in Fig. 2D, and mark changes in the shape of the count of hashtags vs waiting time. At 7 hours we see the start of a shoulder and at around 31 hours the count drops to a very low-level plateau.
(3) The analyzed results in Fig. 3 using 7-31 and 8-30 are similar indicating that the somewhat arbitrary choice of the limits do not influence qualitatively the results, compare Fig. 3 with S2. The proportion of stepwise pattern in the Rome category is slightly higher than the 7-31 division, while the Beauty category remains the same. The division of content categories are also similar.

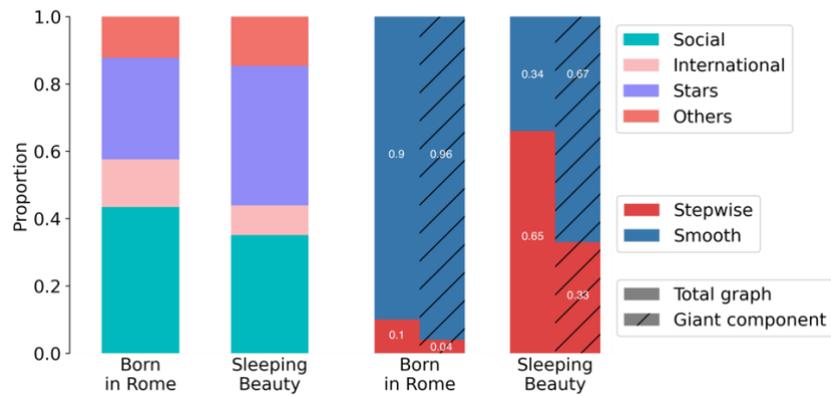

**Supplementary figure S2.** Distribution of hashtags in each categories using 8 hours and 31 hours as division choices of "Born in Rome" and "Sleeping Beauty".

As for the intermediate category between Rome and Beauty, we took a random sample with size 400 from this category and the proportion of stepwise shape is 0.53, which is between the Born in Rome category and the Sleeping Beauty category, as expected.

## SI4. Distribution of sizes of nodes originating HSL hashtags

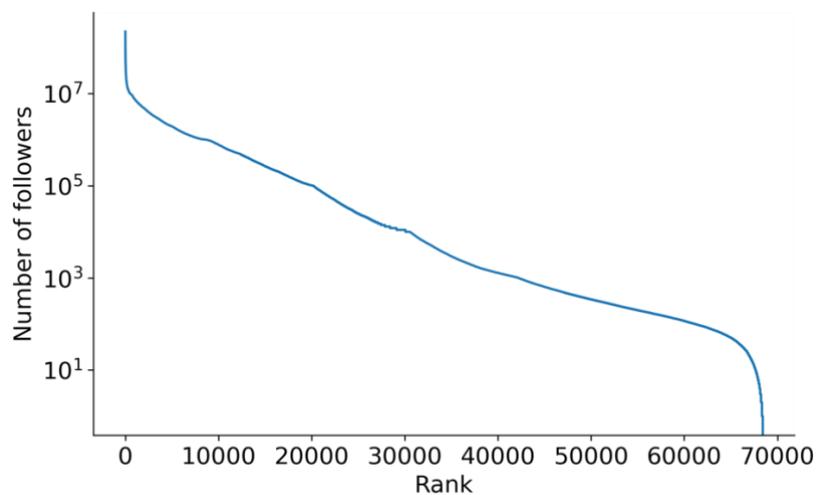

**Supplementary figure S3.** Distribution of number of followers of origin nodes.

As shown in S3, the number of followers of the origin nodes involved in our dataset is ranked in decreasing order. The following table shows the top 20 largest nodes excluding celebrities, which we consider to be the superhubs.

| Rank | Username | Number of followers |
| --- | --- | --- |
| 1 | 超话社区 (Super Talk Community) | 222M |
| 2 | 人民日报 (People's Daily) | 148M |
| 3 | 央视新闻 (CCTV News) | 129M |
| 4 | 新华社 (Xinhua News Agency) | 108M |
| 5 | 头条新闻 (Headline News) | 102M |
| 6 | 新华网 (Xinhuanet) | 95.6M |
| 7 | 人民网 (People's Net) | 82.4M |
| 8 | 新浪新闻 (Sina News) | 77.5M |
| 9 | 中国新闻网 (China News Net) | 75.4M |
| 10 | 中国日报 (China Daily) | 64.9M |
| 11 | 中国新闻周刊 (China News Weekly) | 60.6M |
| 12 | 微博搞笑排行榜 (Weibo Funny List) | 55.9M |
| 13 | 新浪新闻客户端 (Sina News Client) | 52.3M |
| 14 | 环球资讯 (Global Information) | 48.9M |
| 15 | 每日经济新闻 (Daily Economic News) | 48M |
| 16 | 新京报 (Beijing News) | 45.8M |
| 17 | NBA | 42.8M |
| 18 | 新浪娱乐 (Sina Entertainment) | 41.9M |
| 19 | 微博钱包福利 (Weibo Wallet Benefits) | 40.5M |
| 20 | 新闻晨报 (Morning News) | 38.8M |

**Supplementary Table 2.** Top 20 largest nodes (excluding celebrities) on Sina Weibo.

## References

[1] Hao Cui. Evolution of repost networks before getting to the HSL. https://drive.google.com/drive/folders/1JULm8eNswUSOY4PC-cjTtvP4ocJzvaM-



\end{document}